\documentclass[12pt,doublespacing,amsmath,amssym
b,aps]{revtex4}
\usepackage{graphicx}
\begin{document}
\title{{\Large Deuteron photodisintegration with polarized lasers}}
\author{G. Ramachandran$^1$}
\email{gwrvrm@yahoo.com}
\author{S. P. Shilpashree$^{1,2}$}
\email{shilpashreesp@gmail.com}
\affiliation{$^1$GVK Academy, Bangalore, India}
\affiliation{$^2$K. S. Group of Institutions, Bangalore, India}

\begin{abstract}
A model independent theoretical analysis of recent experimental data on deuteron photodisintegration with polarized laser beams is presented. 
We find that it is important to distinguish between  the three isovector E1 amplitudes $E1_v^j$ in reaction channels with total angular 
momentum $j=0,1,2$ and that the isoscalar M1 amplitude $M1_s$is non-zero in the photon energy range $3.5 MeV < E_\gamma < 10 MeV$
\end{abstract}
\maketitle
\noindent \\
Experimental studies \cite{schreiber} have been carried out during the last decade at the Duke Free Electron Laser Laboratory on 
photodisintegration of deuterons using 100$\%$ linearly polarized laser beams from HIGS, in view of the importance of incisive knowledge on 
$d+\gamma\rightleftharpoons n+p$
at the astrophysically relevant range of energies, to sharpen \cite{burles} the predictions of the big bang nucleosynthesis (BBN) and also of 
stellar evolution.\\
\\
The study of  $d+\gamma\rightleftharpoons n+p$ has a long history going back by seven decades to the earliest experimental\cite{chadwick} and 
theoretical\cite{bethe} studies. Traditionally radiative thermal neutron capture has been identified with an isovector $M1_v$ transition, 
while deuteron photodisintegration has been attributed to an isovector $E1_v$ transition. The 10$\%$ discrepancy noted early in the total 
cross section between theory and experiment prompted Breit and Rustgi \cite{breit} to propose a polarized-target-beam test to detect a 
possible isoscalar $M1_s$ transition, but Riska and Brown \cite{riska} explained this with surprising accuracy as due to meson exchange 
currents (MEC). Model calculations taking MEC, isobar current and pair current contributions revealed \cite{nagai} that the dominant $M1_v$ 
transition strength at thermal neutron energies decreases substantially with increasing neutron energy $E_n$, while the $E1$ transition picks 
up the strength in the energy region $10^2 < E_n < 10^3 $ keV and becomes dominant in photodisintegration. This general scenario \cite{nagai} 
is also substantiated by effective field theoretical calculations\cite{ando}. Questions with regard to the less dominant amplitudes have 
however been raised, using different versions of effective field theory \cite{chen} and also using the six quark dressed bag model 
\cite{kukulin}. 
We may recall that model calculations have lead to  traditional forms referred to as Rustgi parametrization\cite{rustgi1} and Partovi 
parametrization \cite{partovi} for the differential cross section. Equivalently the cross section may also be expanded empirically in terms 
of associated Legendre polynomials. For example, the Rustgi parametrization is of the form
\begin{eqnarray}
{d\sigma \over d\Omega}=a+b \sin^2\theta \pm c\cos\theta \pm d \sin^2\theta \cos\theta \pm e \sin^2\theta \cos^2\theta \nonumber \\
+\cos 2\phi[f\sin^2\theta \pm d \sin^2\theta \cos\theta \pm e \sin^2\theta \cos^2\theta]
\end{eqnarray}
where the $\pm $ sign refers to the outgoing protons/neutrons in the c.m. frame. The form used by Schreiber et al \cite{schreiber} is
\begin{equation}
{d\sigma \over d\Omega}={2\pi^2 \over 6} [a+b \sin^2\theta (1+\cos 2\phi)]
\end{equation}
following Weller et al\cite{weller}. A simple analysis using conservation laws reveals all the allowed amplitudes as shown in Table I.
\begin{table}[t]
\caption{ All the allowed multipole amplitudes}
\begin{tabular}{ccc}
\hline
Continuum eigen state \qquad \qquad
& Notation used for the Multipoles\\
\hline
$^1 S_0, I=1$ & $ M1_v $ \\
$^3 S_1, I=0$ & $ M1_s, E2_s$  \\
$^1 P_1, I=0$ & $ E1^{j=0}_s, M2_s $ \\
$^3 P_0, I=1$ & $ E1^{j=0}_v $ \\
$^3 P_1, I=1$ & $ E1^{j=1}_v, M2_v $ 
\\
$^3 P_2, I=1$ & $ E1^{j=2}_v, M2_v, E3_v $ \\
\hline
\end{tabular}
\end{table}
Our model independent approach\cite{sps} for photodisintegration of deuterons using $100\%$ linearly polarized photons leads to the following 
expression
\begin{equation}\label{poldiff}
{d\sigma \over d\Omega}={2\pi^2 \over 6} [a+b \sin^2\theta (1+\cos 2\phi)
-c\cos\theta],
\end{equation}
if all the higher order multipoles are neglected.
The recent experimental data reported by Sawatzky\cite{schreiber} and Blackston \cite{schreiber} have also been analyzed in terms of 
associated Legendre polynomials, where the isotropic term has been normalized to 1. The comparison of this analysis with eq.(3) reveals that 
$c/a$  is non-zero and has values shown in Table II.
\begin{table}[t]
\caption{Estimates of c/a from experiment }
\begin{tabular}{ccc}
\hline
$E_\gamma$ MeV
& ${c \over a}$\\
\hline
3.5 & 0.2325 $\pm$ 0.0722 \\
4 & 0.1084 $\pm$ 0.0391 \\
6 & 0.0160 $\pm$ 0.0141\\
10 & -0.1413 $\pm$ 0.0074\\
14 & -0.056 $\pm$ 0.006 \\
16 & -0.077 $\pm$ 0.006 \\
\hline
\end{tabular}
\end{table} 
It was shown in \cite{sps} for the first time that the $\cos \theta$ term survives, even if we disregard all the higher order multipole 
amplitudes and that the coefficient $c$ in eq(3) is given by
\begin{equation}\label{c}
c = 4\sqrt 6 Re[(2 E1_v^{j=0} + 3 E1_v^{j=1} 
-5 E1_v^{j=2})M1_s^*\big].
\end{equation}
The empirical values presented in Table II shows at once that  
\begin{equation}
M1_s \neq 0
\end{equation}
and that $E1^j_v$ for $j=0,1,2$ can not all be equal.  This is a significant result. \\
\\
It might also be mentioned that we have recently studied \cite{sps2} theoretically the differential cross section for aligned deuterons using 
linearly polarized laser beam. Further work is in progress to distinguish between the three $E1^j_v$ amplitudes, by employing polarized 
deuteron targets which are characterized by both tensor as well as vector polarization. Finally we may mention that it is of crucial interest 
to astrophysics that the interference term $c/a$ between $M1_s$ and $E1^j_v$ amplitudes  shows an increasing trend as $E_\gamma$ decreases 
i.e., as we approach astrophysically relevant energies.

\section*{Acknowledgements}
We are grateful to Professor Blaine Norum for sending us the details of the work of Dr. B. D. Sawatzky and Dr. M. A. Blackston. One of us 
(SPS) is thankful to the Principal and Management of K. S. Group of Institutions for their support and encouragement.


\begin{thebibliography}{}
\bibitem{schreiber} E. C. Schreiber {\it et al.,} Phys. Rev. {\bf C61}, 061604 (2000)
\\ W. Tornow {\it et al.,} Mod. Phys. Lett. {\bf A18}, 282 (2003)
\\ W. Tornow et al, Phys. Lett. {\bf B574}, 8 (2003)
\\ Bradley David Sawatzky, Ph.D Thesis, University of Virginia, (2005) 
\\ M. A. Blackston, Ph. D. Thesis, Duke University (2007)
\\ M. W. Ahmed et al., Phys. Rev. {\bf C77}, 044005 (2008)
\bibitem{burles} S. Burles and D. Tytler, Astrophys. J. {\bf 499}, 699 (1998)\\
S. Burles and D. Tytler, Astrophys. J. {\bf 507}, 732 (1998)
\\ S. Burles, K. M. Nollett, J. W. Truran and M. S. Turner,
Phys. Rev. Lett. {\bf 82}, 4176 (1999)
\bibitem{chadwick}C. Chadwick and M. Goldhaber, Nature (London) {\bf 134}, 237 (1935)
\\ I.E. Amaldi and Fermi, Phys. Rev. {\bf 50}, 899 (1936)
\bibitem{bethe} H. A. Bethe and R. E. Peierls, Proc. Roy. Soc. (London) {\bf A148}, 146 (1935)
\bibitem{breit} G. Breit and M. L. Rustgi, Nucl. Phys. {\bf A161}, 337 (1971)
\bibitem{riska} D. O. Riska and G. E. Brown, Phys. Lett. {\bf B38}, 193 (1972)
\bibitem{nagai} Y. Nagai et. al., Phys. Rev. {\bf 56}, 3173 (1997)
\bibitem{chen} J.- W. Chen, G. Rupak and M. Savage, Phys. Lett. {\bf B464}, 1 (1999)
\\ T.-S. Park, K. Kubodera, D.-D. Min and M. Rho, Phys. Lett. {\bf B472}, 232 (2000)
\bibitem{kukulin} Kukulin et al., Int. J. Mod. Phys. {\bf E11}, 1 (2002)\\
M. M. Kaskulov et al., arXiv nucl-th/0212097
\bibitem{ando} S. Ando et al, arXiv: nucl-th/ 0511074 v1 and references there in.
\bibitem{rustgi1} M. Rustgi, W. Zernik, G. Breit and D. Andrews, Phys. Rev. {\bf 120}, 1881 (1960)
\bibitem{partovi} F. Partovi, Annals of Physics {\bf 27}, 79 (1964)
\bibitem{weller} H.R. Weller {\it et al.,} At.Data. Nucl. Data Tables {\bf 50}, 29 (1992)
\bibitem{sps} G. Ramachandran and S. P. Shilpashree, Phys. Rev. {\bf C74}, 052801 (2006)(R)
\bibitem{sps2} G. Ramachandran, Swarnamala Sirsi and S. P. Shilpashree, Proc. DAE Symp on Nucl. Phys., BITS, Pilani, {\bf 55}, 436 (2010) 
Eds: R. K. Choudhury, D. C. Biswas and K. Mahata  , Board of Research in Nuclear Sciences, DAE, Govt of India
\end{thebibliography}
\end{document}